\documentclass[preprint,showpacs,preprintnumbers,amsmath,amssymb]{revtex4}

\usepackage{graphicx}
\usepackage{dcolumn}
\usepackage{bm}

\begin{document}

\title{\large Transport and spectroscopic properties of\\
superconductor - ferromagnet - superconductor junctions of\\
$La_{1.9}Sr_{0.1}CuO_4$ - $La_{0.67}Ca_{0.33}MnO_3$ - $La_{1.9}Sr_{0.1}CuO_4$}

\author{G. Koren}
\email{gkoren@physics.technion.ac.il} \affiliation{Physics
Department, Technion - Israel Institute of Technology Haifa,
32000, ISRAEL} \homepage{http://physics.technion.ac.il/~gkoren}

\author{T. Kirzhner}
\affiliation{Physics Department, Technion - Israel Institute of
Technology Haifa, 32000, ISRAEL}

\date{\today}
\def\bfig {\begin{figure}[tbhp] \centering}
\def\efig {\end{figure}}

\normalsize \baselineskip=8mm  \vspace{15mm}


\begin{abstract}

Transport and Conductance spectra measurements of ramp-type
junctions made of cuprate superconducting $La_{1.9}Sr_{0.1}CuO_4$ electrodes
and a manganite ferromagnetic $La_{0.67}Ca_{0.33}MnO_3$ barrier are
reported.  At low temperatures below $T_c$, the conductance spectra show Andreev-like broad peaks superposed on a tunneling-like background, and sometimes also sub-gap Andreev resonances. The energy gap values $\Delta$ found from fits of the data ranged mostly between 7-10 mV. As usual, the gap features were suppressed under magnetic fields but revealed the tunneling-like conductance background. After field cycling to 5 or 6 T and back to 0 T, the conductance spectra were always higher than under zero field cooling, reflecting the negative magnetoresistance of the manganite barrier. A signature of superparamagnetism was found in the conductance spectra of junctions with a 12 nm thick LCMO barrier. Observed critical currents with barrier thickness of 12 nm or more, were shown to be an artifact due to incomplete milling of one of the superconducting electrodes.

\end{abstract}

\maketitle

Superconductor - ferromagnet - superconductor (SFS) junctions are interesting since they allow for the investigation of both antagonistic order parameters of S and F by measuring their transport, conductance spectra and behavior under applied magnetic fields. They can also facilitate studies of proximity induced triplet superconductivity (PITS) near inhomogeneities in F by measuring their supercurrents as described in a recent feature article in Physics Today \cite{PT}. Generally, transport in junctions of a superconductor in contact with a normal metal (N) is controlled by the Andreev reflection process \cite{Andreev}, and the penetration of pairs into N is determined by the normal metal coherence length $\xi_N=\sqrt{\hbar D/2\pi k_B T}$ where $D$ is the diffusion coefficient and $T$ is the temperature. If N is replaced by F, the singlet pairs of S can penetrate into F only to a relatively short distance $\xi_F$ ($\xi_F=\hbar v_F/2E_{ex}$ in the clean limit or  $\xi_F=\sqrt{\hbar D/2E_{ex}}$ in the dirty limit) which is affected mostly by the exchange energy $E_{ex}$ rather than the thermal energy $k_BT$ \cite{RMPBergeret,RMPBuzdin}. This is true unless a PITS order is created in F which would lead to longer penetration lengths into F, on the order of $\xi_N$ \cite{Bergeret,Fominov,Eschrig,KB,VE}. Originally, we planned to look for triplet supercurrents in SFS junctions made of the underdoped high temperature superconductor $La_{1.9}Sr_{0.1}CuO_4$ (LSCO10) and the manganite ferromagnet $La_{0.67}Ca_{0.33}MnO_3$ (LCMO). Since all our junctions with barrier thickness of 12 and 20 nm were resistive at 2 K and therefore no PITS could be observed, we decided to focus on the transport, conductance spectroscopy and field behavior of the junctions. We note however, that PITS could be at the origin of previous results where a small supercurrents of 5 $kA/cm^2$ at 4 K had been observed in similar SFS junctions but with $YBa_2Cu_3O_{7-\delta}$ electrodes and a 20 nm thick $La_{0.67}Sr_{0.33}MnO_3$ barrier \cite{Ivanov}. Following our recent study of similar SNS junctions but with a normal $La_{1.65}Sr_{0.35}CuO_4$ barrier where two Andreev-like energy scales were observed \cite{KorenPRL}, in the present study we concentrated on low biases and observed only the lower superconducting energy scale. The negative magnetoresistance of the LCMO barrier at 2 K had also been detected in the conductance spectra of the present work, either under a high magnetic field in the superparamagnetic state of the LCMO barrier or after field cycling.\\

\begin{figure} \hspace{-20mm}
\includegraphics[height=9cm,width=13cm]{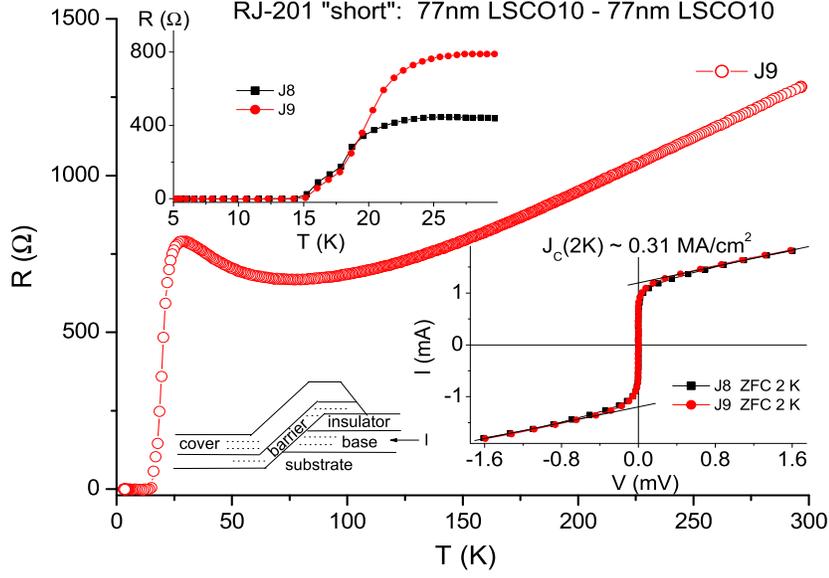}
\vspace{-0mm} \caption{\label{fig:epsart}Resistance versus
temperature of a LSCO10-LSCO10 junction without a barrier (a "short"). The top inset
shows the data of this and another junction on this wafer at low temperature, while the bottom-right inset shows the corresponding I-V curves at 2 K. The bottom-left inset
shows a schematic drawing of a ramp-type junction, where the 77 nm
thick base and cover electrodes are made of LSCO10 and the barrier of thickness $d_F$ is made of LCMO (for the present "shorts" $d_F=0$).}
\end{figure}

Ten ramp-type junctions were prepared as described previously \cite{Nesher123}
along the anti-node direction of the LSCO10 d-wave order parameter
in the geometry shown in the bottom-left inset to Fig. 1, on $10\times 10$ mm$^2$
wafers of (100) $SrTiO_3$ (STO). The different LSCO10 and LCMO layers were grown
epitaxially with the c-axis normal to the wafer, and thus a-b
plane coupling was obtained between the base and cover electrodes.
All junctions had the same structure with $5\,\mu m$ width and 77 nm thick LSCO10 electrodes, but with different thicknesses $d_F$ of the LCMO barrier (0, 12 and 20 nm).
We first prepared junction without a barrier ("short" junctions with $d_F=0$), in order to test the quality and cleanliness of our fabrication process. Typical 4-probe results of the resistance versus temperature are shown in Fig. 1 and its top inset. One can easily see the two distinct transition temperature onsets at 23 and 18 K, which correspond to
the $T_c$ values of the cover and base electrodes, respectively.
The reason for this behavior is that the base electrode on the pristine STO
surface is more strained than the cover electrode which is grown
on a  LCMO layer on top of the ion milled area of the
STO wafer \cite{Locquet,Iguchi}. The current versus voltage (I-V) curves and the critical current density $J_c$ of these junctions at 2 K are shown in the second bottom inset to Fig. 1. The measured $J_c(2K)$ here is 0.31 MA/cm$^2$ which is comparable to the value obtained in the literature for  blanket films (0.4 MA/cm$^2$ at 4.2 K \cite{Iguchi}), and also to that obtained in our own micro bridges patterned on blanket LSCO10 films. This means that our preparation process is quite clean and introduces only a negligible amount of contamination at the interfaces of our junctions. In passing we note that in the supplementary material of a previous publication by our group concerning these short junctions \cite{KorenPRL}, the given $J_c(2K)$ values were in error and the correct values are given here.\\

\begin{figure} \hspace{-20mm}
\includegraphics[height=9cm,width=13cm]{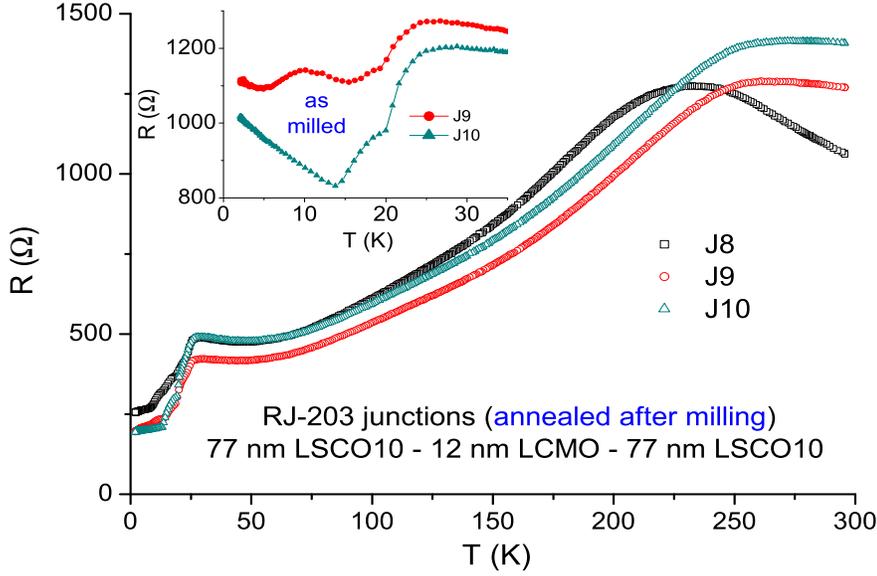}
\vspace{-0mm} \caption{\label{fig:epsart}Resistance versus temperature of three
LSCO10-LCMO-LSCO10 junctions with a $d_F$=12 nm thick barrier. The inset shows the low temperature data of two of these junctions in the as milled state, right after fabrication, but without the further oxygen annealing step as shown in the main panel. }
\end{figure}

Next we present resistance versus temperature results of three junctions with a $d_F$=12 nm thick LCMO barrier (see Fig. 2). The different resistance values at 300 K are due to the different lengths of the normal leads to the junctions. The bending down of the resistance curves of J9 and J10 at 250 K and the peak of J8 at 230 K are due to the transition to ferromagnetism of the LCMO layer at the corresponding $T_{Curie}$ values.  These different $T_{Curie}$ values can be attributed to oxygen inhomogeneities in the LCMO layers. The high sensitivity of the LCMO to oxygen loss is also demonstrated in the inset to Fig. 2 where the low temperature R versus T curves of J9 and J10 are shown right after the fabrication process (the "as milled" state), and before the following oxygen annealing step after which the data of the main panel was taken. In addition to the two transition to superconductivity seen here of the base and cover electrodes as also discussed above concerning the shorts of Fig. 1, one can see that the junctions' resistance below about 10 K ranges between 900 and 1100 $\Omega$ in the as milled state while after the oxygen reannealing it decreased to 200-250 $\Omega$. These junctions however still have a significant resistance at 2 K and therefore have no supercurrent.\\

\begin{figure} \hspace{-20mm}
\includegraphics[height=9cm,width=13cm]{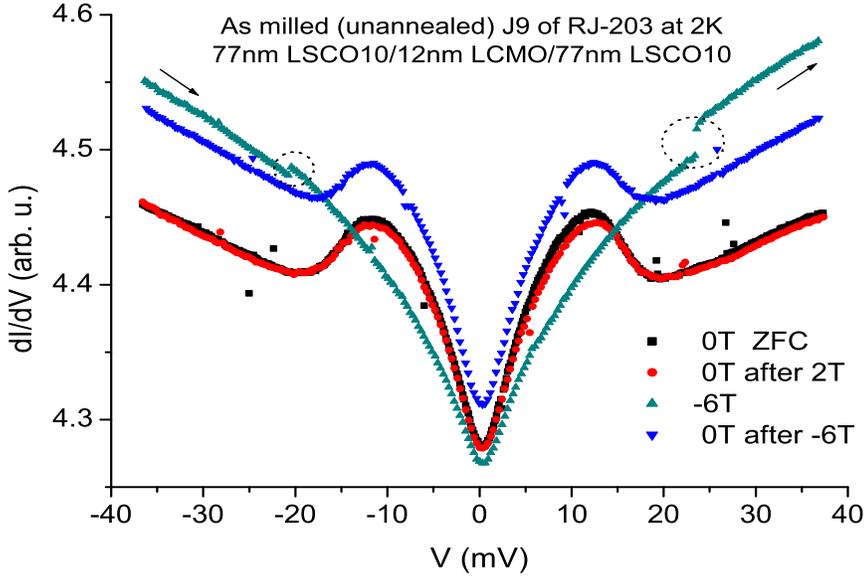}
\vspace{-0mm} \caption{\label{fig:epsart}Conductance spectra of the as milled junction J9 of the inset to Fig. 2, at 2 K and under -6 T and 0 T magnetic fields but with different magnetic field histories.  }
\end{figure}

Fig. 3 shows conductance spectra at 2 K of the as milled junction J9 of the inset to Fig. 2 under zero field with different magnetic histories and under -6 T. The zero field spectra have peaks, similar to coherence peaks in tunneling junctions, at about $\pm$12 mV which are completely suppressed under the -6 T magnetic field. These peaks are therefore due to superconductivity while the spectrum under -6 T can be attributed to a typical background conductance. Hence the observed conductance spectra at zero field $G$(total) seem to originate in a broad Andreev-like feature $G(\Delta)$ superposed on a tunneling-like background conductance $G$(BG). Fig. 4 presents the same zero field cooled (ZFC) data of Fig. 3 normalized to 1 at -20 mV, together with a fit to the BTK model for a d-wave superconductor \cite{Tanaka}. We used $G$(total)=$G(\Delta)+G$(BG) with the fit parameters of the barrier strength Z=0.6, twice the gap energy (for the two interfaces of an SFS junction) $2\Delta$=14.4 mV, the life-time broadening $\Gamma$=1.4 mV and the background conductance 0.95+$|V|^{0.5}$. The fit seems to follow the spectrum quite well and we thus used this fitting procedure throughout this study, but with different background conductance as the case requires. We note however that using a constant background in the fit, yielded a much higher Z value ($\sim$1.5), consistent with tunneling conductance and the high resistance of this junction ($\sim$1 k$\Omega$), but also misfitted the data badly near zero bias and above twice the gap value ($2\Delta$=13.5 mV). Fig. 4 also shows conductance spectra under increasing magnetic fields up to 2 T, where each spectra is normalized by the same normalization factor as used in the ZFC case. The peaks of these spectra are clearly suppressed under increasing magnetic fields, lending further support for their superconductivity origin. In the inset to Fig. 4, the gap values obtained from the corresponding fits of each spectrum, are plotted versus applied field together with an exponential decay fit which seems to fit the data nicely. The factor of 2 (T) in the exponent seems to be related to $H_{c2}$ of the weakened LSCO10 superconductor at the interfaces of the junction.\\

\begin{figure} \hspace{-20mm}
\includegraphics[height=9cm,width=13cm]{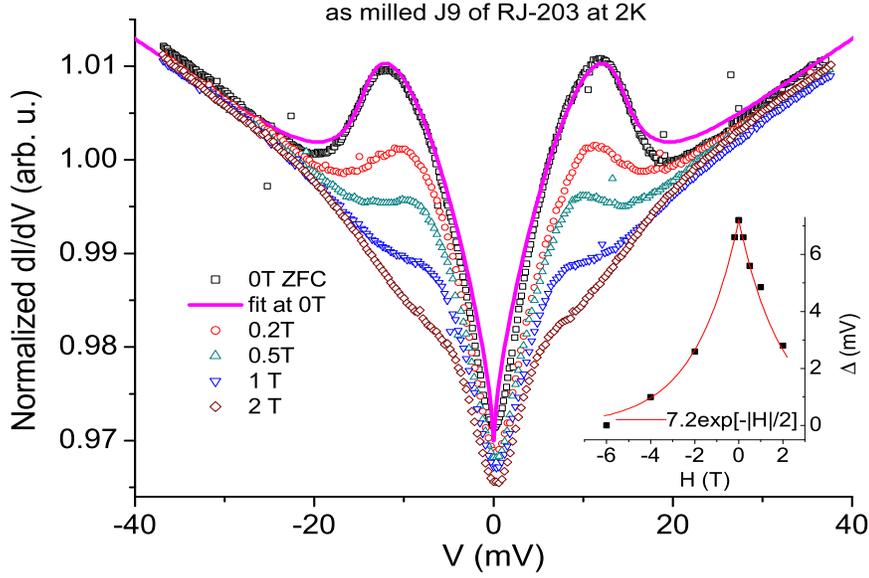}
\vspace{-0mm} \caption{\label{fig:epsart} Normalized conductance spectra of
the same junction as in Fig. 3 at 2 K and under various magnetic fields. A d-wave BTK model fit to the zero field data is shown using Z=0.6, $2\Delta$=14.4 mV, $\Gamma$=1.4 mV and a background of 0.95+0.01$\sqrt{V}$. The inset shows the resulting $\Delta$ values versus field together with an exponential decay fit to the data. }
\end{figure}

We shall now discuss the anomalous magnetic hysteresis effects observed in Fig. 3. Field cycling at 2 K after ZFC to 2 T and back to 0 T, yields a conductance spectrum which almost overlaps the ZFC spectrum, except for a slight tendency for decreasing conductance in the region the peaks. This behavior could  be explained by a small magnetic flux which is trapped in the LSCO10 electrodes and suppresses superconductivity like that seen in Fig. 4 under an external field. However, after field cycling to -6 T and back to 0 T, exactly the opposite effect is found where the conductance is much enhanced compared to the ZFC case. To understand this behavior, we look first at the conductance spectrum of Fig. 3 under -6 T. In the bias regime where $|V|\leq$15 mV, the conductance is lower under -6 T than in the ZFC case, as in Fig. 4 with lower fields. At higher biases though, the conductance under -6 T is significantly higher than that of the ZFC spectrum, and this seems to originate in the negative magnetoresistance of the LCMO ferromagnet. Further support for this interpretation comes from the observation of jumps (or steps) of increasing conductance in the spectrum under -6 T (see in the dotted circles). We note that the conductance data in our experiments was measured by ramping the voltage slowly, in about 2-3 min., from negative to positive values (see the arrows in Fig. 3). During this relatively long recording time, occasional conductance jumps occurred, but only at high fields and quite rarely. These jumps can be attributed to increased magnetization of the LCMO layer by magnetic domain flips, thus increasing the conductance due to the negative magnetoresistance of this manganite material. We point out here that the polarization of a similar manganite ($La_{0.67}Sr_{0.33}MnO_3$) is 78$\pm$4 \% \cite{Soulen} while the coercive field of LCMO is about 0.05 T \cite{Ziese}. Thus one would expect that already under 2 T all the domains of the LCMO layer will be fully oriented along the field. This is apparently not the case, as unnegligible magnetization changes occur at higher fields, which lead to the enhanced conductance even after going back to zero field.\\

\begin{figure} \hspace{-20mm}
\includegraphics[height=9cm,width=13cm]{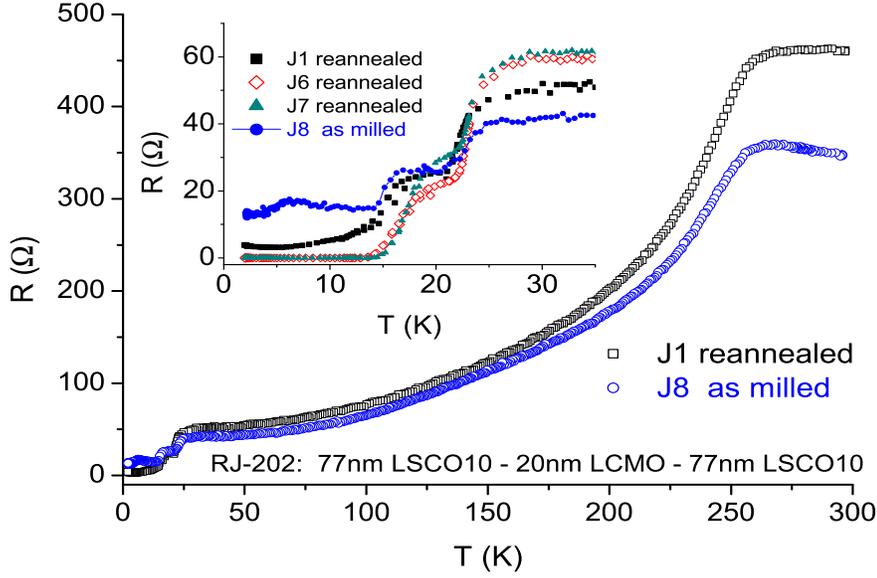}
\vspace{-0mm} \caption{\label{fig:epsart} Resistance versus temperature of two LSCO10 - 20 nm LCMO - LSCO10 junctions, J8 in the as milled state and J1 after oxygen reannealing. The inset shows a zoom up on low temperatures with data on two more reannealed junctions on the same wafer that show zero resistance below 14 K. }
\end{figure}

\begin{figure} \hspace{-20mm}
\includegraphics[height=9cm,width=13cm]{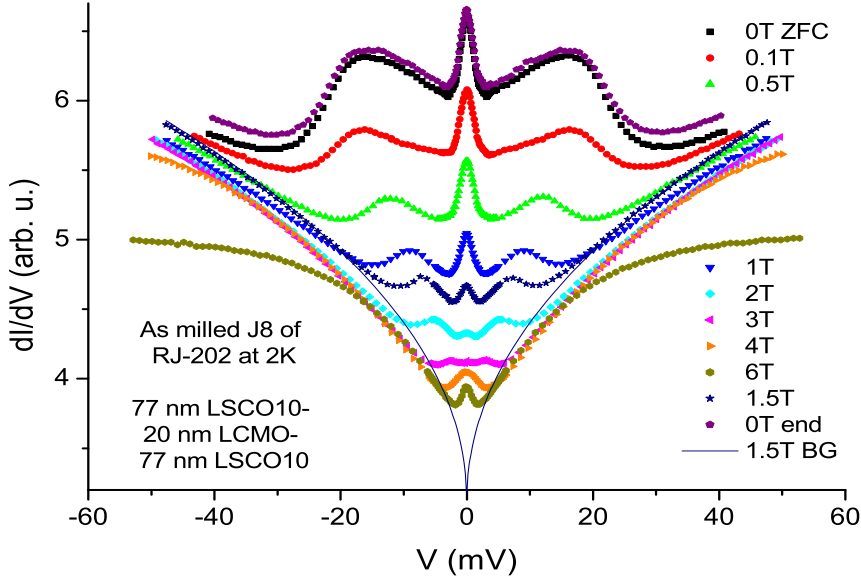}
\vspace{-0mm} \caption{\label{fig:epsart}Conductance spectra at 2 K of
the as milled junction J8 of Fig. 5 under various magnetic fields. Also shown by the solid line is the background tunneling-like conductance curve which is a power low fit to the 1.5 T data at biases above about 20 mV and below -20 mV. This yields a dI/dV background curve of 3.12+0.42$\times |V^{0.485}|$. }
\end{figure}

Fig. 5 and its inset show resistance versus temperature results of junctions with a 20 nm thick LCMO barrier in the as milled state and after oxygen reannealing. Due to the thicker LCMO layer, $T_{Curie}$ here is around 260-270 K, which is a bit higher than in Fig. 2 with the 12 nm thick LCMO layer. The inset shows again that the resistance below 10 K of the as milled junction is higher than that of the oxygen annealed ones where some of them have even zero resistance (which is puzzling, and will be explained later on). Fig. 6 presents conductance spectra at 2 K of the as milled junction J8 of Fig. 5 under increasing fields from 0 T under ZFC to 6 T and back to 1.5 T and 0 T. The ZFC spectrum has a typically broad Andreev peak which is clearer than in Fig. 3 but has the same origin. There is also a narrow zero bias conductance peak (ZBCP) now which is a result of bound states in the LSCO10 d-wave superconductor \cite{Tanaka,Tal,KorenLevy}. As observed in Fig. 3, we also find here that after the 6 T field cycling, the spectrum at 0 T has a higher conductance than that of the ZFC one. This effect can also be seen above 40 mV bias in Fig. 6, where under the returning 1.5 T field, the conductance is higher than under the 1 T and even 0.5 T field. This result shows that the conductance enhancement due to the negative magnetoresistance of LCMO is a robust effect. We also note that the ZBCP and gap features seems to merge at 6 T. The basic background conductance for our fits was obtained by fitting for instance the 1.5 T data below -20 mV and above 20 mV to a power low of $A+B\times |V|^q$ where $A$, $B$ and $q$ are the fitting parameters. The result is shown by the solid curve in Fig. 6. Fine tuning of these parameters was later made for each specific fit to get the best fit to the data. \\

 Next we discuss the opposite behaviors of the conductance in Figs. 3 and 6 under a 6 T field and at high bias of $|V|\geq 20$ mV, where the observed conductance is higher or lower than the ZFC conductance, respectively. Fig. 6 also shows that at 6 T, the conductance increase versus bias below -20 mV and above 20 mV is much slower as compared to that under other fields in this figure. The main differences between the two junctions of Figs. 3 and 6 are the different barrier thickness (12 and 20 nm) as well as the junctions' resistance at zero bias (1110 and 13 $\Omega$ at 2 K). While the resistance at 2 K of the junction with the 20 nm thick barrier (J8 of Figs. 5 and 6) agrees quite well with the value calculated from the junction geometry and LCMO resistivity (R$\sim$10 $\Omega$ using $\rho$=20 m$\Omega$cm for the as milled layer), the resistance of the junction with the 12 nm barrier thickness (J9 of Figs. 2-4) is way too high compared to the calculated value. One would expect first that this is a result of the highly strained state of the thinner LCMO layer which is sandwiched in between the two LSCO10 electrodes. However, since the current has to cross the two strained interfaces in both type of junctions (with the 12 and 20 nm thick LCMO barriers), strains alone can not explain the large differences of resistance and magnetoresistance between Figs. 3 and 6, and another explanation is needed. It was found by  Colino  and de Andres that at low temperatures the resistance and magnetoresistance  in LCMO films thinner than about 10 nm are very large, orders of magnitude larger than in 15 and 30 nm thick films \cite{Colino}. They attributed this unique behavior of the ultra thin films to superparamagnetism \cite{Blamire,Aarts}, where the conductance is controlled by the phase difference between adjacent magnetic domains. The conductance can easily be blocked at zero or low fields by an anti-phase boundary, but can be open again by aligning the domains in-phase under high fields. Our data in Figs. 3 and 6 is therefore consistent with this phenomenon, where the 12 nm thick film is close to the superparamagnetic state, while the 20 nm thick film behaves as a normal ferromagnetic metal.\\

\begin{figure} \hspace{-20mm}
\includegraphics[height=9cm,width=13cm]{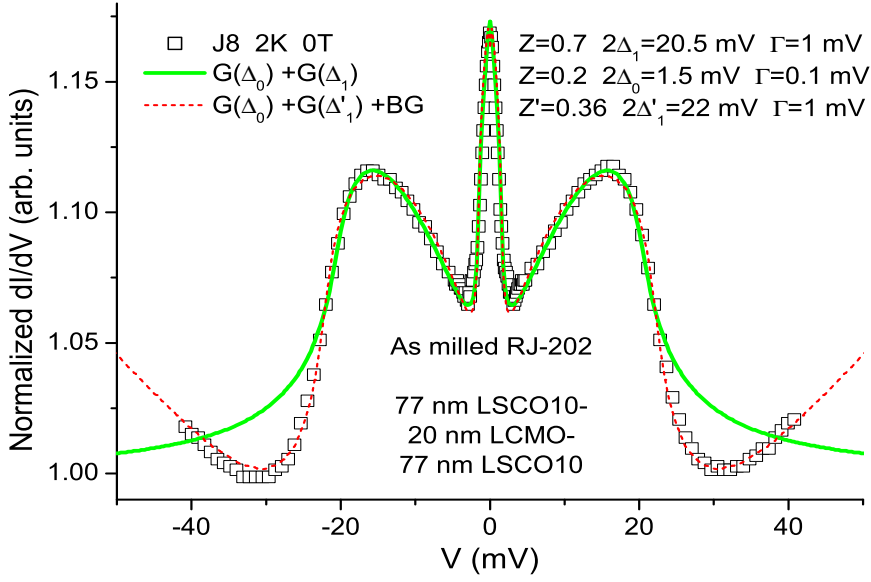}
\vspace{-0mm} \caption{\label{fig:epsart}Conductance spectrum of
the as milled junction J8 of Fig. 6 at 2 K after ZFC with two fits to
the d-wave BTK model. One fit with a constant background (solid line) and the other with a background of 0.71+0.03$\times V^{0.6}$ (dashed line). }
\end{figure}

Fig. 7 shows the normalized conductance spectrum of J8 of Fig. 6 at 2 K after ZFC with two fits to the BTK model for a d-wave superconductor \cite{Tanaka}, one with a  normal background conductance (a constant) and the other with the tunneling-like background conductance (the power low fit) as shown in Fig. 6. In the fits we used $G$(total)=$G(\Delta_0)+G(\Delta_1)+G$(BG) where the three terms represent the ZBCP, Andreev gap, and background contributions to the conductance, respectively. With the constant background, a larger barrier strength Z of 0.7 had to be used compared to the Z'=0.36 which was used with the tunneling-like background fit. Apart from this, the resulting Andreev gap values of $\Delta_1$=10.25 mV and $\Delta'_1$=11 mV are quite similar, but larger than the  $\Delta_1$=7.2 mV value obtained from the fit of Fig. 4. The clear difference between the two fits is that one misfits above the gap while the other fits all regions nicely. So as long as only the $\Delta_1$ value is required, one can do with the normal background fit which necessitates less free parameters.\\

\begin{figure} \hspace{-20mm}
\includegraphics[height=9cm,width=13cm]{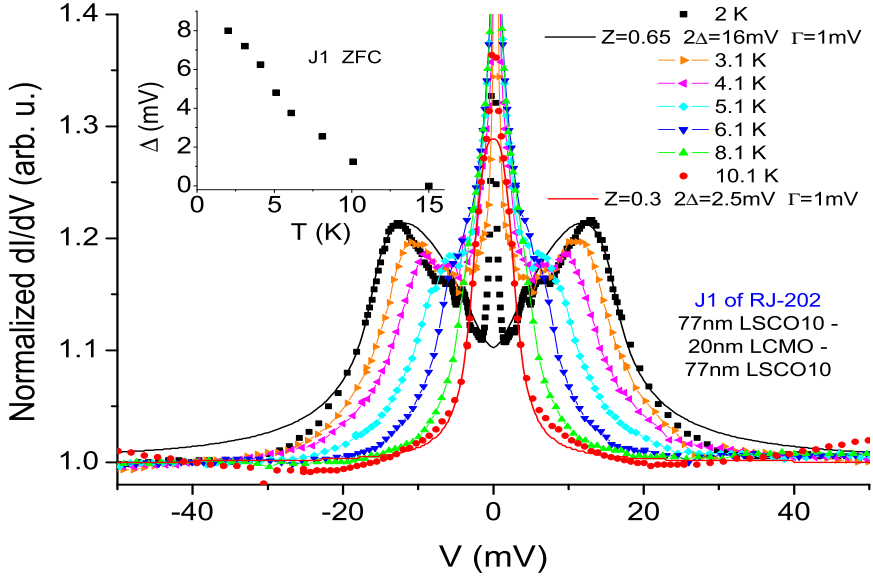}
\vspace{-0mm} \caption{\label{fig:epsart}Normalized conductance spectra of
the reannealed junction J1 of Fig. 5 at various temperatures. Fits to the d-wave BTK model with a constant background are shown for 2 and 10.1 K. The inset shows the gap energy $\Delta$ determined by the fits versus T. }
\end{figure}

Fig. 8 presents normalized conductance spectra of the reannealed junction J1 of Fig. 5 at various temperatures together with fits at 2 and 10.1 K. The ZFC spectrum at 2 K is similar to that of Fig. 7, but with a more robust ZBCP and also with a series of sub-gap resonances which will be analyzed in Fig. 9. The fit of this spectrum is done disregarding the ZBCP and resonances while using a normal background. Similar fits were carried out for all the other spectra in this figure and the resulting energy gap values $\Delta$ are plotted in the inset versus temperature. The $\Delta$ value at 2 K is 8 mV now which is in between the previous results of Figs. 4 and 7 (7-10 mV), and comparable to the scanning tunneling spectroscopy result $\Delta_1$=9.5 mV \cite{Yuli}. The $\Delta$ versus T result of the inset has clearly a sub-BCS behavior which is typical of the cuprates with the d-wave order parameter. Fig. 9 shows a zoom up on the ZFC spectrum of Fig. 8 at 2 K, where the series of dips are marked with arrows. These are due to sub-gap Andreev resonances which occur at bias values of $V_n=2\Delta/n$ where $n$ is a natural number \cite{Nesher123}. In the inset, these $V_n$ values are plotted versus $1/n$ for the positive and negative bias dips, and the line is a linear fit of this data. From this fit $\Delta$=7.1 mV is obtained, which is very close to the $\Delta$=7.2 mV value found in Fig. 4.\\

\begin{figure} \hspace{-20mm}
\includegraphics[height=9cm,width=13cm]{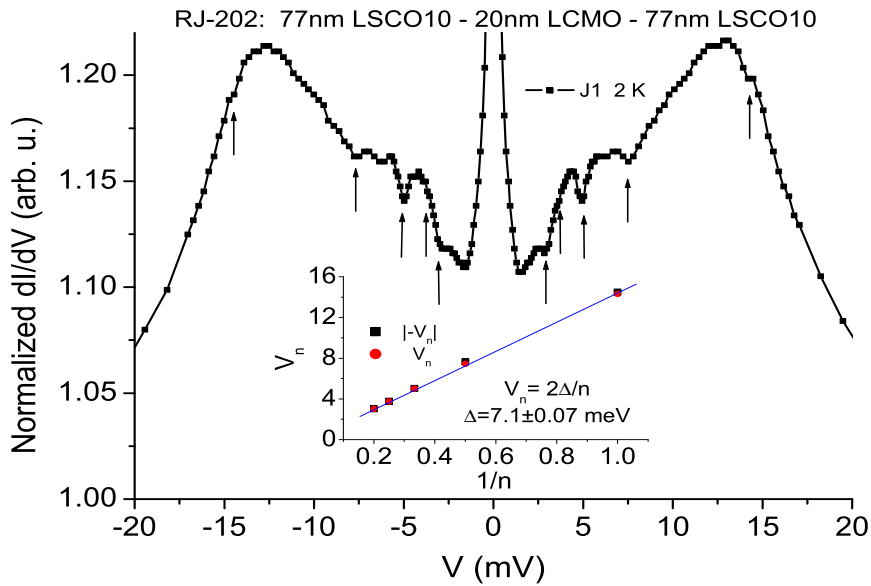}
\vspace{-0mm} \caption{\label{fig:epsart}A zoom up on the low bias ZFC data at 2 K of Fig. 8. The voltage values $V_n$ of the n$^{th}$ Andreev resonance dip in the conductance spectrum are shown by the arrows and plotted in the inset as a function of 1/n. The linear fit to the $|-V_n|$ and $V_n$ series versus 1/n yields an energy gap value $\Delta$ of 7.1 mV. }
\end{figure}

Finally, we shall explain the origin of the puzzling zero resistance observed in the reannealed junctions J6 and J7 of the inset to Fig. 5 below 14 K which turned out to be an artifact. First, since these junctions with a 20 nm thick LCMO barrier (RJ-202) were prepared before the junctions with the 12 nm thick barrier (RJ-203), we thought that a proximity induced triplet superconductivity (PITS) is involved, as these junctions had high supercurrent densities at 2 K on the order of 2 MA/cm$^2$. Then in retrospect this is absurd, in view of the significantly resistive junctions found at 2 K with the thinner 12 nm thick barrier (see Fig. 2) and also the much lower $J_c(2K)$=0.31 MA/cm$^2$ of the short junctions (see bottom inset to Fig. 1). We suspected that these high supercurrents are due to superconducting shorts, and if so, an additional milling step without any resist mask would have remove these conjectured cover electrode shorts. Fig. 10  and its inset show the R versus T results after this addition step where a 10 nm top layer was milled away from the wafer and further oxygen reannealing was performed. One can see that all junctions are resistive now at 2 K with no trace of supercurrents as the corresponding I-V curves are linear (not shown). Also seen in the main panel is that some $T_{Curie}$ are suppressed to 170-210 K, which is due to oxygen inhomogeneities again that might have formed during the last milling step. We thus conclude that no supercurrents neither PITS are observed in the present junctions. Generally, people would not publish an artifact as reported here. The reason that we do, is to warn the readers of possible similar problems when a very long range proximity or PITS effects are observed.\\

\begin{figure} \hspace{-20mm}
\includegraphics[height=9cm,width=13cm]{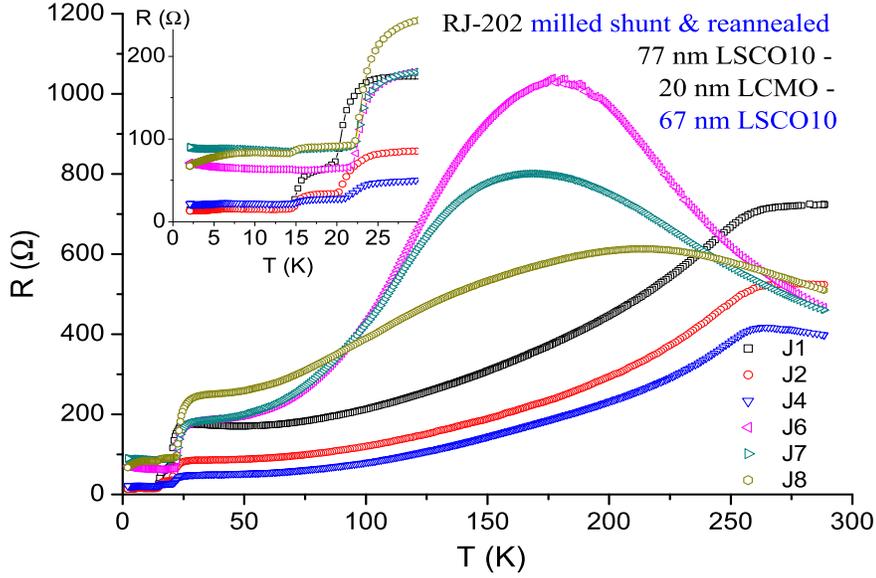}
\vspace{-0mm} \caption{\label{fig:epsart}Resistance versus temperature of six junctions on the wafer of Fig. 5 which were further milled (without a resist) to remove  a 10 nm layer from the top of the cover electrode, and further reannealed in oxygen after that. The zero resistance observed below 14 K in the inset to Fig. 5 is now gone, and all the junctions are resistive at low temperatures as can clearly be seen in the inset.  }
\end{figure}

To summarize, conductance spectroscopy of LSCO10-LCMO-LSCO10 SFS junctions show prominent broad Andreev peaks and sub-gap Andreev resonance series from which the energy gap values of LSCO10 were deduced. Anomalous hysteretic effects were observed after field cycling which were attributed to the negative magnetoresistance of the LCMO barrier. We also found a signature of superparamagnetism in the conductance spectra of the junctions with the 12 nm thick LCMO barrier. And finally, we presented an artifact from which people may learn more than they can expect.\\

{\em Acknowledgments:}  This research was supported in part by the
Israel Science Foundation, the joint German-Israeli DIP project and the Karl Stoll Chair in advanced materials at the Technion.\\

\bibliography{AndDepBib.bib}

\bibliography{apssamp}

\end{document}